\documentstyle[12pt]{article}
\input{epsfig.sty}
\newcommand{\beq}{\begin{eqnarray}}
\newcommand{\eeq}{\end{eqnarray}}
\begin{document}
\title{Charm, Bottom, Strange Baryon Masses Using QCD Sum Rules}
\author{Leonard S. Kisslinger and Bijit Singha\\
Department of Physics, Carnegie Mellon University, Pittsburgh, PA 15213}
\date{}
\maketitle
\begin{abstract}
We use the method of QCD Sum Rules to estimate the masses of the charm 
baryon, $\Lambda^+_c$, bottom baryon $\Lambda^0_b$, strange baryon $\Lambda^0_s$
and compare them to their experimental values.

\end{abstract}


\noindent
PACS Indices: 12.15.Ji, 13.30.Eg, 12.38.Lg, 11.50.Li

\section{Introduction}

  The method of QCD sum rules were introduced by Shifman, Vainshtein, and 
Zakharov \cite{SVZ} to estimate properties of hadrons using 2-point correlators.
See L.J. Reinders, et.al.\cite{RRY85} for a detailed description of this
method for estimating various properties of hadrons. We use this method for
estimating the masses of the charm baryon $\Lambda^+_c$, bottom baryon 
$\Lambda^0_b$, and strange baryon $\Lambda^0_s$. Since the contribution
from the quark condensates\cite{aer12} and the gluon condensate\cite{bv97} is 
negligible for the 2-point correlator needed to estimate the mass of the 
$\Lambda^+_c$, $\Lambda^0_b$, $\Lambda^0_s$ we only use the main diagram, shown 
in Figure 1 for $\Lambda^+_c$ in the following section. The experimental values 
of the masses are found in Ref\cite{PDG16}. After the figures we review 
previous publications using QCD sum rules to estimate baryon and heavy quark 
masses.
\section {QCD Sum Rules for the mass of the $\Lambda^+_c$}
  The two-point correlator in momentum space is used to estimate the mass of 
the $\Lambda^+_c$, a charmed baryon. The two-point correlator is 
\beq
\label{2-ptcorrelator}
\Pi_2(p) &=& i\int d^4x \; e^{ip \cdot x} <0 \mid T[\eta_{\Lambda^+_c}(x)\,
 \bar{\eta}_{\Lambda^+_c}(0)] \mid 0>,
\eeq 
where $\eta_{\Lambda^+_c}$ is the current for $\Lambda^+_c$:
 
\beq
\label{eta}
\eta_{\Lambda^+_c}  &=& \epsilon^{efg} [u^{eT} C \gamma_\mu d^f] \gamma^5 
\gamma^\mu c^g \; ,
\eeq
with $e,f,g$ color indices and $\epsilon^{efg}$ resulting in $\eta_{\Lambda^+_c}$
having color=0, $u,d$ are up and down quarks and $c$ a charm quark.
\newpage
The two-point correlator for $\Lambda^+_c$ is illustrated in Figure 1

\begin{figure}[ht]
\begin{center}
\epsfig{file=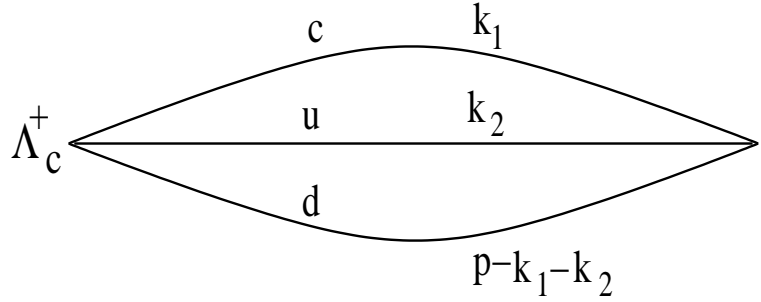,height= 4cm,width=10cm}
\end{center}
\caption{The $\Lambda^+_c$ (udc)}
\end{figure}

Using the methods of QCD Sum Rules\cite{RRY85} the $T[\eta_{\Lambda^+_c}(x)\,
 \bar{\eta}_{\Lambda^+_c}(0)]$ is replaced by the Trace
\beq
\label{2-ptcorrelator-trace}
\Pi_2(p) &=& i\int \frac{d^4k_1}{(2\pi)^4}\frac{d^4k_2}{(2\pi)^4}  
Tr[S_c(k_1)\gamma^5 \gamma^\mu S_u(k_2)\gamma^5 \gamma^\mu S_d(p-k_1-k_2)] \,
\eeq 
where $S_c,S_u,S_d$ are the charm, up, and down quark propagators.
With $\not{p} = p_\mu \gamma^\mu$
\beq
\label{S(k} 
      S_c(p)&=& i \frac{\not{p}+M_c}{p^2 - M_c^2} \, 
\eeq
where $M_c\simeq 1.27$ GeV is the mass of the charm quark. $S_u,S_d$ have 
similar expressions with $m_u \simeq m_d \equiv m \ll M_c$. 
From Eqs(\ref{2-ptcorrelator-trace},
\ref{S(k})

\beq
\label{2-ptcorrelatorz}
\Pi_2(p) &=& \int \frac{d^4k_1}{(2\pi)^4}\frac{d^4k_2}{(2\pi)^4} 
\frac{Tr[(\not{k}_1+M_c)\gamma^5 \gamma^\mu (\not{k}_2+m)\gamma^5 \gamma^\mu
((\not{p}-\not{k}_1-\not{k}_2)+m)]}{[(k_1^2 - M_c^2)(k_2^2 - m^2)
((p-k_1-k_2 )^2 - m^2)]} 
\eeq 

Carrying out the trace one obtains
\beq
\label{2-ptcorrelator-final}
\Pi_2(p)&=& 8\int \frac{d^4k_1}{(2\pi)^4}\frac{d^4k_2}{(2\pi)^4} 
\frac{2M_cm^2 +M_ck_2\cdot (p-k_1-k_2)+2mk_1\cdot(p-k_1-k_2)+mk_1\cdot k_2}
{[(k_1^2 - M_c^2)(k_2^2 - m^2)((p-k_1-k_2 )^2 - m^2)]} \nonumber \\
&\equiv&8\int \frac{d^4k_1}{(2\pi)^4 (k_1^2 - M_c^2)}\Pi_{k_1}(p) \ .
\eeq 

 Defining $\bar{p}=p-k_1, k=k_2$, 
\beq
\label{2-ptcorrelator-k}
\Pi_{k_1}(p) &=& \int \frac{d^4k}{(2\pi)^4} 
\frac{2M_c m^2 +M_ck\cdot (\bar{p}-k)+2mk_1\cdot(\bar{p}-k)+ mk_1\cdot k}
{[(k^2 - m^2)((\bar{p}-k )^2 -m^2)]} \ .
\eeq
Defining
\beq
\label{pi1}
\Pi_1(p)&=&\int \frac{d^4k}{(2\pi)^4}\frac{1}{[(k^2 - m^2)((\bar{p}-k )^2-m^2)]}
 \ ,
\eeq 
and using $1/(k^2-m^2)= \int_0^\infty d\alpha e^{-\alpha(k^2-m^2)}$ ,
$D=4-\epsilon$, $\bar{k}=k-\beta\bar{p}/(\alpha +\beta)$, 
$\int\frac{d^D\bar{k}}{(2\pi)^4} e^{-(\alpha+\beta)\bar{k}^2} =
1/(4\pi(\alpha+\beta))^{D/2}$, and
$\alpha\rightarrow \rho \alpha, \beta\rightarrow \rho \beta, \delta(\rho-\alpha
-\beta) = \delta(1-\alpha-\beta)/\rho$,  one obtains for $\Pi_1(p)$ 

\newpage
\beq
\label{correlator1}
\Pi_1(p) &=&  \int_0^\infty d\rho \int_0^1 d\alpha  \int_0^1 d\beta 
\frac{1}{(4 \pi)^{D/2}}\rho^{1-D/2} e^{-\rho [-m^2 + \alpha(1-\alpha)\bar{p}^2]}
\delta(1-\alpha-\beta) \ .
\eeq

  Integrating Eq(\ref{correlator1}) by parts one obtains

\beq
\label{Pi-k(p)}
\Pi_1(p) &=& \frac{1}{(4 \pi)^2}(2m^2-\bar{p}^2/2)I_0(\bar{p}) {\rm with} \\
  I_0(p) &=&  \int_0^1 d\alpha \frac{1}{-m^2 +\alpha(1-\alpha) p^2} \nonumber
 \ .
\eeq

From Eq(\ref{2-ptcorrelator-k}) we also need
\beq
\label{p1mu}
\Pi_1^\mu(p)&=& \int \frac{d^4k}{(2\pi)^4}\frac{k^\mu}{[(k^2-m^2)((\bar{p}-k )^2
-m^2)]}=\frac{\bar{p}^\mu}{(4 \pi)^2}[(m^2-\bar{p}^2/4)I_0(\bar{p})-\frac{7}{4}]
\ , 
\eeq

Therefore from Eqs(\ref{2-ptcorrelator-k},\ref{Pi-k(p)},\ref{p1mu}),
 
\beq
\label{k1correlator}
\Pi_{k_1}(p)&=&\frac{1}{(4\pi)^2}[(2m k_1\cdot \bar{p}+m^2M_c)(2m^2-\bar{p}^2/2)
I_0(\bar{p}) \\
&&+(M_c\bar{p}^2-mk_1\cdot \bar{p})((m^2-\bar{p}^2/4)I_0(\bar{p})-\frac{7}{4})]
 \nonumber \ .
\eeq

From Eqs(\ref{2-ptcorrelator-final},\ref{k1correlator}), dropping the 7/4 term
which vanishes with a Borel transform

\beq
\label{2-ptcorrelator-end}
\Pi_2(p)&=&\frac{8}{(4\pi)^4}\int_0^1 d\alpha
(2m^4M_cI_1(p)+3m^3I_2(p)+\frac{3m^2M_c}{2}I_3(p)\nonumber \\
 && -\frac{3 m}{4}I_4(p) -\frac{M_c}{4}I_5(p)) \ .
\eeq

With  $k=k_1$
\beq
\label{I_1}
 I_1(p)&=& \int\frac{d^4k}{(2\pi)^4}\frac{1}{(k^2-M_c^2)
[\alpha(1-\alpha)(p-k)^2-m^2]} \\
 I_2(p)&=& \int\frac{d^4k}{(2\pi)^4}\frac{k\cdot(p-k)}{(k^2-M_c^2)
[\alpha(1-\alpha)(p-k)^2-m^2]}  \\
 I_3(p)&=&  \int\frac{d^4k}{(2\pi)^4}\frac{(p-k)^2}{(k^2-M_c^2)
[\alpha(1-\alpha)(p-k)^2-m^2]} \\
 I_4(p)&=&  \int\frac{d^4k}{(2\pi)^4}\frac{k\cdot(p-k)(p-k)^2}{(k^2-M_c^2)
[\alpha(1-\alpha)(p-k)^2-m^2]} \\
 I_5(p)&=& \int\frac{d^4k}{(2\pi)^4}\frac{(p-k)^2(p-k)^2}{(k^2-M_c^2)
[\alpha(1-\alpha)(p-k)^2-m^2]}  \ .
\eeq

\newpage

  Dropping terms that vanish with a Borel transform one can show that
\beq
\label{I345}
 I_3(p)&=& \frac{m^2}{\alpha(1-\alpha)} I_1(p) \nonumber \\
 I_4(p)&=& \frac{m^2}{\alpha(1-\alpha)}  I_2(p) \nonumber \\
 I_5(p)&=& \frac{m^4}{\alpha^2(1-\alpha)^2} I_1(p) \, .
\eeq

  Using $1/[\alpha(1-\alpha)(p-k)^2-m^2]=\int_0^\infty d\lambda e^{-\lambda
[\alpha(1-\alpha)(p-k)^2-m^2]}$, $1/(k^2-M_c^2)= 
 \int_0^\infty d\kappa e^{-\kappa(k^2-M_c^2)}$
\begin{eqnarray}
I_1(p) &=& \int \frac{d^4 k}{(2 \pi)^4}\int_0^\infty d\kappa d \lambda~ 
e^{- \kappa(k^2 - M_c^2) - \lambda[-m^2 + \alpha(1 - \alpha)(p - k)^2]}
\end{eqnarray}
 
  Carrying out the momentum integral with $d^4 k\rightarrow d^Dk, 
D\equiv 4-\epsilon$, using $\kappa \rightarrow \rho \kappa$, $\lambda 
\rightarrow \rho \lambda$,$\delta(\rho-\kappa-\lambda)=\delta(1-\kappa-
\lambda)/\rho$; and carrying out
the $\lambda$ integral, and using  $\kappa \rightarrow \rho \kappa$, 
$\lambda \rightarrow \rho \lambda$, $\delta(\rho - \kappa - \lambda) = 
\delta(1- \kappa - \lambda)/\rho$ one obtains
\beq
I_1(p) &=& \frac{1}{(4 \pi)^2}\int_0^\infty d \rho \int_0^1 d \kappa~
\rho^{(1 - D/2)} \frac{1}{[\kappa + \alpha(1 - \alpha)(1 - \kappa)]^{D/2}} ~
 e^{ - \rho a} 
\eeq
with
\beq
a &=& \frac{\kappa(1 - \kappa)\alpha (1 - \alpha)p^2}{\kappa + 
\alpha(1 - \alpha)(1- \kappa)} - \kappa M_c^2 - (1 - \kappa)m^2
\eeq

 Carrying out $\int_0^\infty d \rho$ and using $a^{- \frac{\epsilon}{2}} = 
e^{-(\epsilon/2) \ln a} = 1 - (\epsilon/2) \ln a$
\beq
I_1(p) &=& \frac{1}{(4 \pi)^2} \Gamma \left(\frac{\epsilon}{2}\right) 
\int_0^1 d\kappa a^{-\frac{\epsilon}{2}} \frac{1}{[\kappa + \alpha(1 - \alpha)
(1 - \kappa)]^2} \nonumber \\
&=& \frac{1}{(4 \pi)^2} \int_0^1 d \kappa 
\frac{1}{[\kappa + \alpha(1 - \alpha)(1 - \kappa)]^2} \times (- \ln a ) 
\; .
\eeq

 Evaluating $I_2(p)$ in a similar way one finds
\begin{eqnarray}
I_2(p) &=& \int \frac{d^4 k}{(2 \pi)^4}
\frac{k.(p - k)}{(k^2 - M_c^2)\left[ -m^2 + \alpha(1-\alpha)(p - k)^2 \right]} \\
&=& p_\mu \int \frac{d^4 k}{(2 \pi)^4}
\frac{k^\mu}{(k^2 - M_c^2)\left[ -m^2 + \alpha(1-\alpha)(p - k)^2 \right]} 
\nonumber\\
&& - \int \frac{d^4 k}{(2 \pi)^4}
\frac{M_c^2}{(k^2 - M_c^2)\left[ -m^2 + \alpha(1-\alpha)(p - k)^2 \right]} \\
&=& p_\mu \bar{\Pi}_1^\mu (p) - M_c^2 I_1(p) \; ,
\end{eqnarray}
with
\begin{eqnarray}
\bar{\Pi}_1^\mu (p) &=& \int \frac{d^4 k}{(2 \pi)^4} ~k^\mu
\int_0^\infty d\kappa d \lambda~ e^{
- \kappa(k^2 - M_c^2) - \lambda[-m^2 + \alpha(1 - \alpha)(p - k)^2]}
 \nonumber\\
&=& \int_0^\infty
d \kappa d \lambda 
\int \frac{d^4 k}{(2 \pi)^4} ~ k^\mu ~\exp \Big[ -\big( \kappa + \lambda 
\alpha (1 - \alpha)  \big) \left( k - \frac{\lambda\alpha(1 - \alpha)p}
{\kappa + \lambda\alpha(1 - \alpha)} \right)^2
 \nonumber\\
&& \qquad - \lambda \alpha (1 - \alpha)p^2 + \frac{(\lambda\alpha
(1 - \alpha)p)^2}{\kappa + \lambda\alpha(1 - \alpha)} + (\kappa M_c^2 + 
\lambda m^2) \Big] \; . 
\end{eqnarray}
\newpage

  Using $k^\mu \rightarrow k^\mu + \frac{\lambda\alpha(1 - \alpha)p^\mu}{\kappa +
 \lambda\alpha(1 - \alpha)}$, one finds

\begin{eqnarray}
\bar{\Pi}_1^\mu (p) &=& 
\int_0^\infty d \kappa d \lambda 
\int \frac{d^4 k}{(2 \pi)^4} ~ \left[k + \frac{\lambda\alpha(1 - \alpha)p}
{\kappa + \lambda\alpha(1 - \alpha)} \right]^\mu 
\times \exp \Big[ -\left( \kappa + \lambda \alpha (1 - \alpha)  \right) k^2 
\nonumber\\
&& \qquad - \lambda \alpha (1 - \alpha)p^2 + \frac{(\lambda\alpha
(1 - \alpha)p)^2}{\kappa + \lambda\alpha(1 - \alpha)} + (\kappa M_c^2 + \lambda
 m^2) \Big] \; .
\end{eqnarray}

  Integrating over the four-momenta
\begin{eqnarray}
\label{PiBarPInLn}
\bar{\Pi}_1^\mu (p) &=& \frac{p^\mu}{(4 \pi)^2}\int_0^1 d \kappa
\frac{\alpha(1 - \alpha)(1 - \kappa)}{[ \kappa + \alpha (1 - \alpha) 
(1 - \kappa)]^3} \times ( - \ln a) \,; 
\end{eqnarray}
and therefore
\begin{eqnarray}
I_2(p) &=& p^2 \bar{\Pi}(p) - M_c^2 I_1(p) \; ,
\end{eqnarray}
where 
\begin{eqnarray}
\bar{\Pi}(p) &=& \frac{1}{(4 \pi)^2} \int_0^1 d \kappa \frac{\alpha(1 - \alpha)
(1 - \kappa)}{[ \kappa + \alpha (1 - \alpha) (1 - \kappa)]^3} \times ( - \ln a)
\; .
\end{eqnarray}

Therefore from Eqs(\ref{2-ptcorrelator-end},\ref{I345})
\beq
\label{2-ptcorrelator-finalz}
\Pi_2(p) &=&  \frac{8}{(4 \pi)^4}\int_0^1 d \alpha ~[2 m^4 M_c I_1(p) + 3 m^3 
\left\{p^2\bar{\Pi}(p) - M_c^2 I_1(p)\right\} + \frac{3 m^4 M_c}{2 \alpha 
(1 - \alpha)} I_1 (p) \nonumber\\
&& -\frac{3 m^3}{4 \alpha (1 - \alpha)} \left\{ p^2 \bar{\Pi}(p) - M_c^2 I_1 (p)
 \right\} - \frac{M_c m^4}{4 \alpha^2 (1 - \alpha)^2}I_1(p)] \nonumber\\
&=&  \frac{8}{(4 \pi)^4}
\int_0^1 d \alpha ~\Big[\left\{2 m^4 M_c - 3 m^3 M_c^2 + \frac{3M_c m^4}
{2 \alpha (1 - \alpha)}  + \frac{3 m^3 M_c^2}{4 \alpha (1 - \alpha)} - 
\frac{M_c m^4}{4 \alpha^2 (1 - \alpha)^2} \right\} I_1 (p) \nonumber\\
&& + \left\{3 m^3 - \frac{3 m^3}{4 \alpha (1 - \alpha)}
\right\} p^2 \bar{\Pi}(p)\Big] \ .
\eeq

  A Borel transform $\mathcal{B}$ of $\Pi_2(p)$ gives
\beq
\label{Pi2mB}
\Pi_2 (M_B) &=& \frac{8}{(4 \pi)^4}\int_0^1 d \alpha ~\Big[\left\{
2 m^4 M_c - 3 m^3 M_c^2 + \frac{3M_c m^4}{2 \alpha (1 - \alpha)}  + \frac{3 m^3
M_c^2}{4 \alpha (1 - \alpha)} - \frac{M_c m^4}{4 \alpha^2 (1 - \alpha)^2}
\right\} {\mathcal{B}} [I_1 (p)] \nonumber\\
&& + \left\{3 m^3 - \frac{3 m^3}{4 \alpha (1 - \alpha)}\right\} {\mathcal{B}} 
[p^2 \bar{\Pi}(p)] \Big]
\eeq

Note that
\beq
\label{bln}
{\mathcal{B}}[\ln (p^2 - b^2)] &=& - M_B^2~ e^{- b^2/M_B^2}, \nonumber \\
{\mathcal{B}}[p^2 ~\ln (p^2 - b^2)] &=& - M_B^2(b^2 - M_B^2)~ e^{- b^2/M_B^2}.
\eeq

\newpage

  From Eqs(\ref{Pi2mB},\ref{bln})
\beq
\label{Pi2mbfinal}
\Pi_2 (M_B) &=& - \frac{8}{(4 \pi)^6}\int_0^1 d \alpha \int_0^1 d \kappa 
\left(2 m^4 M_c - 3 m^3 M_c^2 + \frac{3M_c m^4}{2 \alpha (1 - \alpha)} 
+ \frac{3 m^3 M_c^2}{4 \alpha (1 - \alpha)}
- \frac{M_c m^4}{4 \alpha^2 (1 - \alpha)^2} \right) \nonumber\\ 
&& \times \left(\frac{- M_B^2 ~  e^{ - b^2/M_B^2 }}{[\kappa + 
\alpha (1 - \alpha)(1 - \kappa)]^2} \right) \\
&& - \frac{8}{(4 \pi)^6}\int_0^1 d \alpha \int_0^1 d \kappa 
\left(3 m^3 - \frac{3 m^3}{4 \alpha (1 - \alpha)}
\right) \left(\frac{-M_B^2 (b^2 - M_B^2)~\alpha(1 - \alpha)(1 - \kappa) 
e^{ - b^2/M_B^2 }}{[\kappa + \alpha (1 - \alpha)(1 - \kappa)]^3} \right) 
\nonumber \; .
\eeq

It is convenient for $\Pi_2 (M_B)$ to use
\begin{eqnarray}
\label{pi2mbz}
\Pi_2 (M_B) &=& \frac{8 M_B^2}{(4 \pi)^6}\int_0^1 d \alpha~ d \kappa~
\frac{g(\alpha, \kappa) M_5(\alpha) + 3 m^3 M_2(\alpha, \kappa)}
{g(\alpha, \kappa)^3}
~ e^{- b(\alpha, \kappa)^2/M_B^2} \; ,
\end{eqnarray}
where

\begin{eqnarray}
\label{pi2mbzz}
g(\alpha, \kappa) &=& \kappa + \alpha (1 - \alpha)(1 - \kappa),  \\
b(\alpha, \kappa)^2 &=& g(\alpha, \kappa) \Big[
\frac{M_c^2}{\alpha(1 - \alpha)(1 - \kappa)} + 
\frac{m^2}{ \kappa \alpha (1 - \alpha)}
\Big], \nonumber\\
M_2(\alpha, \kappa) &=& \Big( 1 - \frac{1}{4 \alpha(1 - \alpha)} \Big)
\Big[b(\alpha, \kappa)^2 - M_B^2 \Big], \nonumber \\
M_5( \alpha) &=& 2 m^4 M_c - 3 m^3 M_c^2 + \frac{3M_c m^4}{2 \alpha (1 - \alpha)}
 + \frac{3 m^3 M_c^2}{4 \alpha (1 - \alpha)} - \frac{M_c m^4}{4 \alpha^2 
(1 - \alpha)^2} \nonumber \; .
\end{eqnarray}

  From Eqs(\ref{pi2mbz},\ref{pi2mbzz}), dropping the factor of 
$\frac{8}{(4 \pi)^6}$ as only the shape of $\Pi_2 (M_B)$ is needed to
estimate the mass, as in  Ref\cite{kc15} , obtains the result 
shown in Figure 2.

\begin{figure}[ht]
\begin{center}  
\epsfig{file=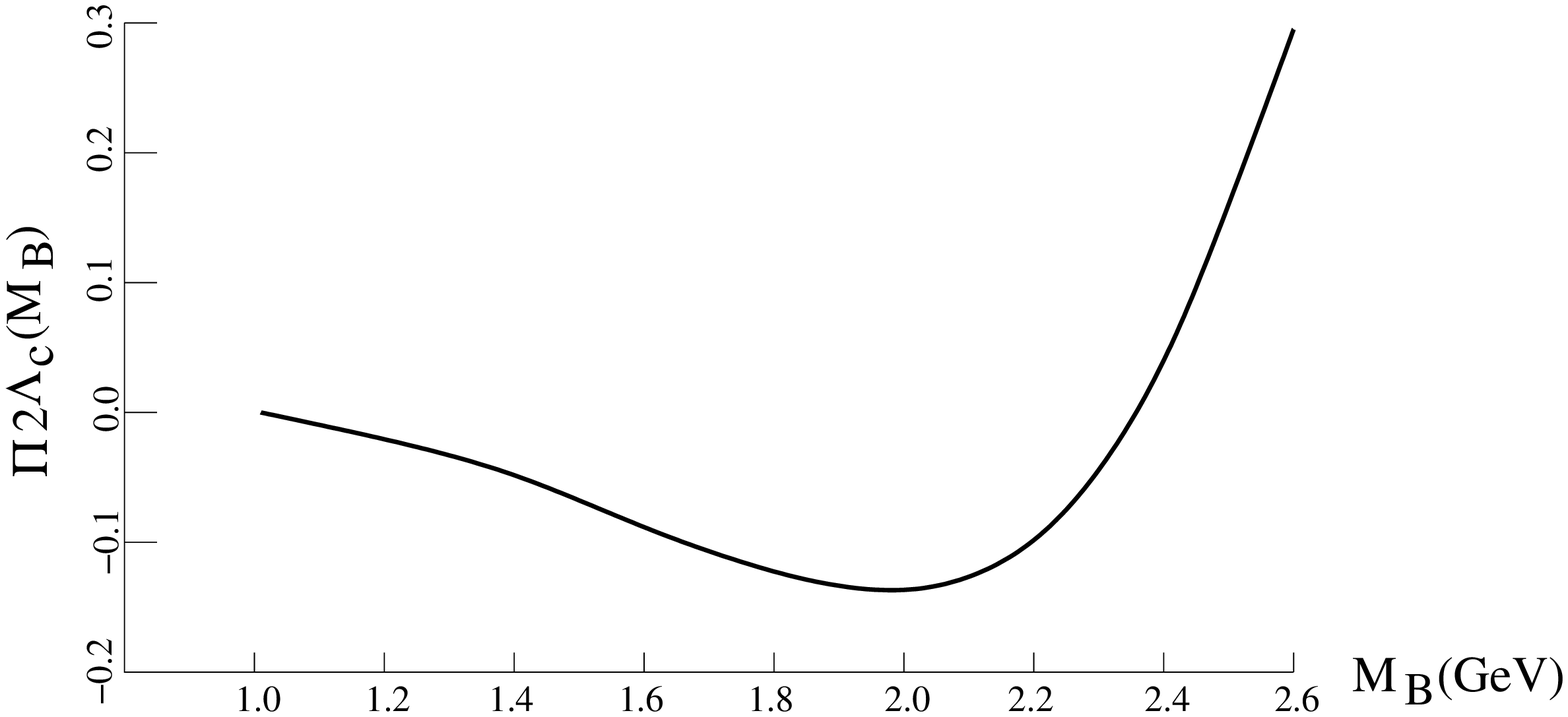,height=6 cm,width=12cm}
\caption{Two-point correlator $\Pi_2$ for $\Lambda^+_c$  as a function of the 
Borel mass $M_B$}
\label{}
\end{center}
\end{figure}
\newpage
\section {QCD Sum Rules for the masses of the $\Lambda^0_b$
and $\Lambda^0_s$}
The two point correlator illustrated in Figure 1 and derived in 
Eqs(\ref{2-ptcorrelator}-\ref{2-ptcorrelator-final}) for $\Lambda^+_c$
are the same for $\Lambda^0_b$,  $\Lambda^0_s$ with  quarks $c\rightarrow b$, 
$c\rightarrow s$ in Eqs(\ref{2-ptcorrelator},\ref{eta}) and the figure; and 
$M_c\rightarrow M_b$, $M_c\rightarrow M_s$ in 
Eqs(\ref{2-ptcorrelator-trace}-\ref{2-ptcorrelator-final}).

  The calculation of the masses of $\Lambda^0_b$ and $\Lambda^0_s$ use
the equations Eqs(\ref{pi2mbz},\ref{pi2mbzz}) with $M_c\simeq 1.27$ GeV
replaced by $M_b \simeq$ 4.18 GeV for the mass of $\Lambda^0_b$  and 
$M_s \simeq$ 96 MeV for the mass of $\Lambda^0_s$.

  The results for $M_{\Lambda^0_b}$ and $M_{\Lambda^0_s}$ are shown in Figures 3 and 
4. 

\begin{figure}[ht]
\begin{center}  
\epsfig{file=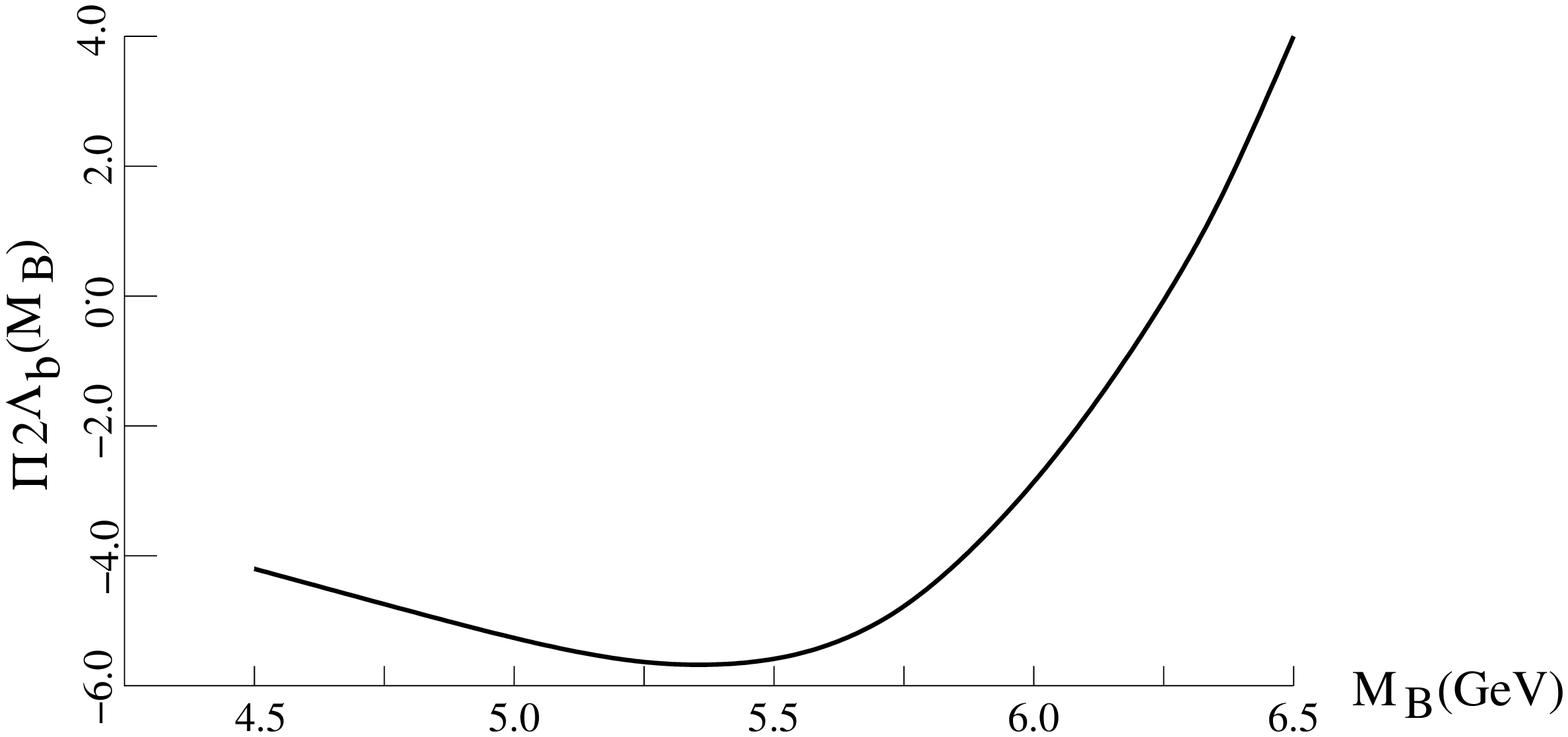,height=6 cm,width=12cm}
\caption{Two-point correlator $\Pi_2$ for $\Lambda^0_b$  as a function of the 
Borel mass $M_B$}
\label{}
\end{center}
\end{figure}

\begin{figure}[ht]
\begin{center}  
\epsfig{file=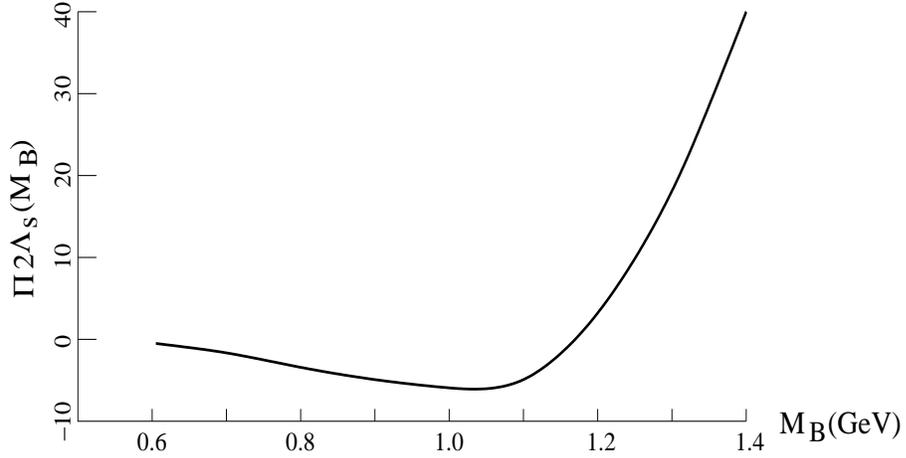,height= 6 cm,width=12cm}
\caption{Two-point correlator $\Pi_2$ for $\Lambda^0_s$  as a function of the 
Borel mass $M_B$}
\label{}
\end{center}
\end{figure}
\newpage

  More than three decades ago sum rules analogous to those of SVZ\cite{SVZ}
were used\cite{Ioffe81} to estimate the mass difference between the 
$\Sigma(3/2^+)$ and the $\Delta(3/2^+)$. Using the strange quark 
mass $m_s$ =150 MeV, Ioffe estimated that $M_{\Sigma}-M_{\Delta} \simeq 125$ MeV, 
while current experimental results\cite{PDG16} are $M_{\Sigma}-M_{\Delta} \simeq
 -130$  MeV. Since the  curent value of $m_s \simeq 96$ MeV\cite{PDG16}, this 
could be an explanation of Ioffe's result. More recently Ioffe\cite{Ioffe08} 
gave evidence that the violation of chiral symmetry in the QCD vacuum is the
origin of baryon masses, and estimated $M_{\Delta} = 1.30 \pm 0.18$ GeV
while current experimental result\cite{PDG16} is $M_{\Delta}\simeq 1.51$ GeV.

Also, using QCD spectral sum rules\cite{bcdn} attempts to estimate the masses 
of the charm baryon $\Lambda^+_c$ and bottom baryon $\Lambda^0_b$ were made, 
but the authors were not able to make reliable predictions for the 
$\Lambda^+_c,\Lambda^0_b$ masses.

  Sum rules have also been used to estimate the masses of quarks. Using a
Pseudoscalar Sum Rule\cite{ck06} the strange quark mass was estimated to be
$m_s \simeq 105 \pm 6 \pm 7$ MeV while the current value\cite{PDG16} is
$m_s \simeq 96$ MeV. Using QCD sum rules\cite{dgp94} the charm quark mass
was estimated as $m_c = 1.46 \pm 0.07$ GeV while the current value\cite{PDG16} 
is $m_c = 1.27 \pm 0.03$ GeV. Using QCD spectral sum rules\cite{nar94} Narison
found the mass of the bottom quark $m_b = 4.23^{+.03}_{-.04}$ GeV while the current
value\cite{PDG16} is $m_b = 4.18^{+.04}_{-.03}$ GeV.
 
\section{Results and Conclusions}

  Using the method of QCD Sum Rules the mass of $\Lambda^+_c$, $\Lambda^0_b$,
and $\Lambda^0_s$ were estimated. The masses and theoretical uncertainty
are estimated using minimum and spread near the minimum in the plot of 
$\Pi_2 (M_B)$  in Figures 2, 3, and 4.

From Figure 2 $M_{\Lambda^+_c} =2.01 \pm 0.3$ GeV compared to the experimental 
value\cite{PDG16} 2.29 GeV. From Figure 3 $M_{\Lambda^0_b} =5.34 \pm 0.25$ GeV 
compared to the experimental value\cite{PDG16} 5.62 GeV. From Figure 4 
$M_{\Lambda^0_s} =1.05 \pm 0.1$ GeV compared to the experimental 
value\cite{PDG16} 1.116 GeV. 

We conclude from our results that the method of QCD Sum Rules\cite{SVZ,RRY85} 
can be used to estimate the masses of charm, bottom, and strange baryons.

\vspace{1cm} 
\Large{{\bf Acknowledgements}}\\
\normalsize
This work was supported in part by a grant from the Pittsburgh Foundation
and in part by the Carnegie Mellon University Department of Physics.

\end{document}